# RGB-Thermal Infrared Fusion for Robust Depth Estimation in Complex Environments

Zelin Meng and Takanori Fukao

***Abstract*—Depth estimation in complex real-world scenarios remains a challenging task, particularly when relying on a single modality such as visible light or thermal infrared (THR) imagery. This paper introduces a novel multimodal depth estimation framework that improves accuracy and robustness by leveraging the complementary characteristics of RGB and THR data. The RGB modality offers rich texture and color cues, while the THR modality captures thermal signatures, providing resilience under adverse lighting conditions, including extreme illumination and low-light environments. The proposed model incorporates a distinctive fusion strategy, EGFusion, which comprises two key components: the Mutual Complementary Attention (MCA) module for cross-modal feature alignment, and the Edge Saliency Enhancement Module (ESEM) for preserving fine edge details. Extensive experiments conducted on the MS2 and ViViD++ datasets show that our model consistently generates high-quality depth maps across a range of challenging environments, such as nighttime scenes, heavy rain, and high-glare conditions. The robust estimation results and real-time performance demonstrate the model's strong potential for real-world applications that demand reliable depth perception, including autonomous driving, robotics, and augmented reality.**

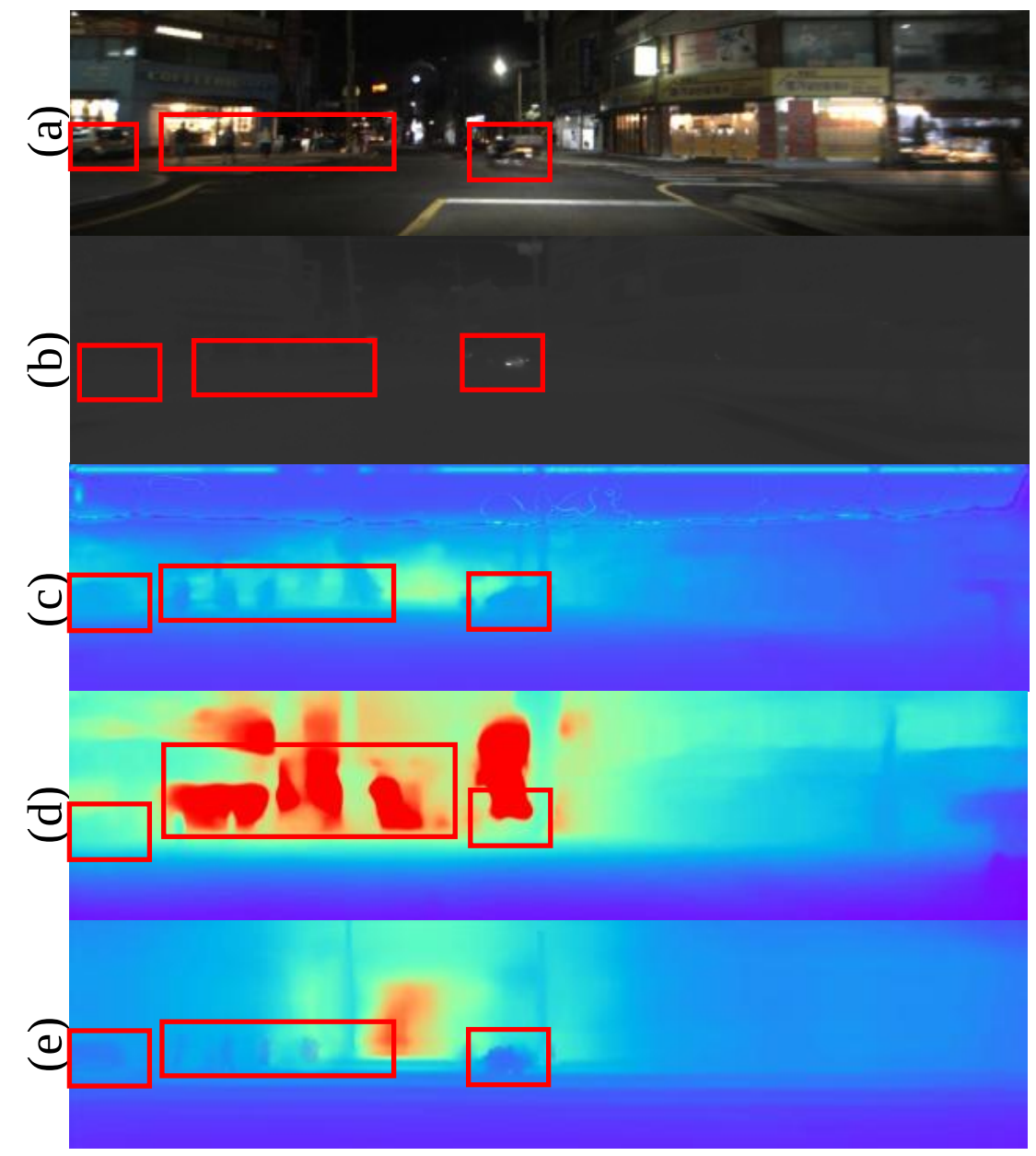

Figure 1. Predicted depth maps from different estimation Methods. (a): RGB input. (b): THR input. (c) Prediction from RGB-based method. (d): Prediction from THR-based method. (e): Prediction from RGB-THR fusion-based method.

## I. INTRODUCTION

Robust depth estimation is an essential yet challenging task in environment perception, which plays a critical role in applications like robotics, and intelligent transportation systems [1]. Precise depth perception is vital for scene comprehension, obstacle detection, and navigation. However, in real-world settings—especially under difficult conditions such as nighttime, heavy rain, fog, or intense glare—depth estimation techniques often face difficulties in maintaining accuracy and robustness [2, 3].

Conventional methods depend on single-modal inputs, mainly RGB images or thermal infrared (THR) imagery. RGB-based approaches utilize rich texture and color details, making them effective in well-lit environments [4, 5]. However, their performance declines significantly in low-light conditions or when visibility is hindered. On the other hand, THR-based methods detect heat signatures, allowing for reliable perception in darkness or adverse weather [6, 7].

Despite these benefits, THR images lack fine texture details and high spatial resolution, resulting in less accurate depth estimations, especially in geometrically complex scenes or low-contrast areas [8].

Although both RGB and THR modalities show strong potential, their independent effectiveness is often limited under challenging environmental conditions. To address this problem, multimodal fusion has emerged as a promising solution that leverages the complementary strengths of RGB and THR data. RGB images provide detailed structural and color information, while THR images enhance the performance in low-visibility scenarios by detecting thermal patterns [9, 10]. By integrating these two modalities, multimodal depth estimation models can achieve higher accuracy and better generalization in a wide range of environmental settings [11].

Despite significant progress, multimodal depth estimation still faces several unresolved challenges [12, 13]. For instance, the intrinsic discrepancies between RGB and THR data,

The author is with the Graduate School of Ixxxxxxxx Department of xxxxx The University of xxxx

including differences in spatial resolution and noise properties, and information distribution, making effective fusion difficult [14, 15]. Secondly, accurate depth estimation in complex environments demands a seamless integration of global semantic context and detailed local features [15, 16]. Thirdly, it remains a major challenge to design multimodal fusion strategies that are scalable, adaptable, and capable of generalizing well under complex environment conditions [17, 18].

In this study, we propose a novel multimodal depth estimation framework designed to address the above-mentioned challenges. Our approach incorporates a Mutual Complementary Attention (MCA) mechanism and an Edge Saliency Enhancement Module (ESEM) to facilitate more effective fusion of RGB and thermal (THR) features. By leveraging the complementary advantages of both modalities, the proposed model delivers robust and accurate depth estimation across a wide range of environmental conditions. Furthermore, an edge refinement module is introduced to improve depth prediction, particularly around object boundaries and in texture-sparse regions, thereby enhancing overall performance in complex real-world scenarios. The main contributions of this work are summarized as follows:

1. We propose a dual-branch feature extraction architecture specifically designed for RGB and thermal (THR) inputs, allowing for independent and modality-specific representation learning. The RGB branch is responsible for capturing fine-grained textures and high-level semantic context, while the THR branch is optimized for thermal radiation perception, providing a more accurate representation of temperature distribution within the scene.
2. To fully exploit the complementary characteristics of RGB and thermal (THR) modalities, we propose an Edge-Guided Fusion (EGFusion) module that integrates a Mutual Complementary Attention (MCA) mechanism for cross-modal feature alignment and an Edge Saliency Enhancement Module (ESEM) for structure-aware information fusion. The MCA employs a Query-Key-Value framework to dynamically model inter-modal feature relationships, thereby facilitating more effective cross-modal interactions and improving depth estimation in scenarios with severe occlusion or complex lighting conditions. Simultaneously, the ESEM enhances prediction accuracy along object boundaries and in low-contrast regions by leveraging the inherent temperature contrast provided by THR data.
3. We conduct extensive experiments on the MS2 dataset, where our model demonstrates competitive performance compared to state-of-the-art multimodal depth estimation methods. Both quantitative evaluations and ablation studies confirm the robustness and effectiveness of our approach, particularly under challenging conditions such as low illumination, occlusion, and low-contrast environments.

## II. Related Works

Depth estimation methods can be broadly categorized into single-modality and multimodal approaches. Recent advancements in deep learning and self-supervised learning have significantly improved depth estimation accuracy, particularly in challenging environments [19].

### *A. Depth Estimation from RGB*

Monocular depth *estimation* from RGB images has been extensively studied in computer vision. Traditional methods rely on handcrafted features and structure-from-motion techniques, while recent advances leverage deep learning to extract depth information from single images. Vision transformers (ViTs) have been introduced for dense prediction tasks, replacing convolutional networks to improve depth estimation accuracy by utilizing global receptive fields. Similarly, large models like "Depth Anything" focus on scaling up the dataset to improve generalization and zero-shot capabilities, significantly enhancing depth estimation performance on public benchmarks [12]. Laplacian pyramid-based methods have also been proposed to refine depth estimation by preserving multi-scale depth residuals, improving both boundary sharpness and global consistency [16]. Self-supervised approaches further mitigate the need for labeled depth data by employing photometric consistency constraints, leading to state-of-the-art results in various datasets, including KITTI and NYUv2 [9, 25].

RGB-based depth estimation utilizes rich texture and color information but suffers from sensitivity to lighting variations and occlusions [25, 28]. In contrast, THR-based methods leverage thermal signatures for robustness in low-light conditions but lack spatial resolution and texture details. Recent advances, including unsupervised domain adaptation and self-supervised learning, aim to enhance single-modality depth estimation by improving generalization and reducing domain shifts [26].

### *B. Depth Estimation from THR*

Conversely, thermal infrared (THR)-based methods provide a unique advantage by detecting heat signatures, ensuring robustness against lighting variations. These approaches are especially effective in environments with limited visible light, such as nighttime or foggy conditions. Recent advancements have introduced techniques for estimating depth and ego-motion from thermal images by utilizing multi-spectral consistency losses [7, 22, 23, 24]. Additionally, adversarial multi-spectral adaptation leverages unpaired RGB-THR images to enhance feature extraction and depth estimation for thermal inputs [20]. Some works propose direct depth estimation from single thermal images by incorporating generative adversarial networks (GANs) and contour-aware constraints, which help mitigate the lack of texture in thermal images. Large-scale multi-spectral stereo datasets have also been created to benchmark depth estimation models under various lighting and weather conditions, aiding in the development of more robust thermal-based depth estimation techniques [27].

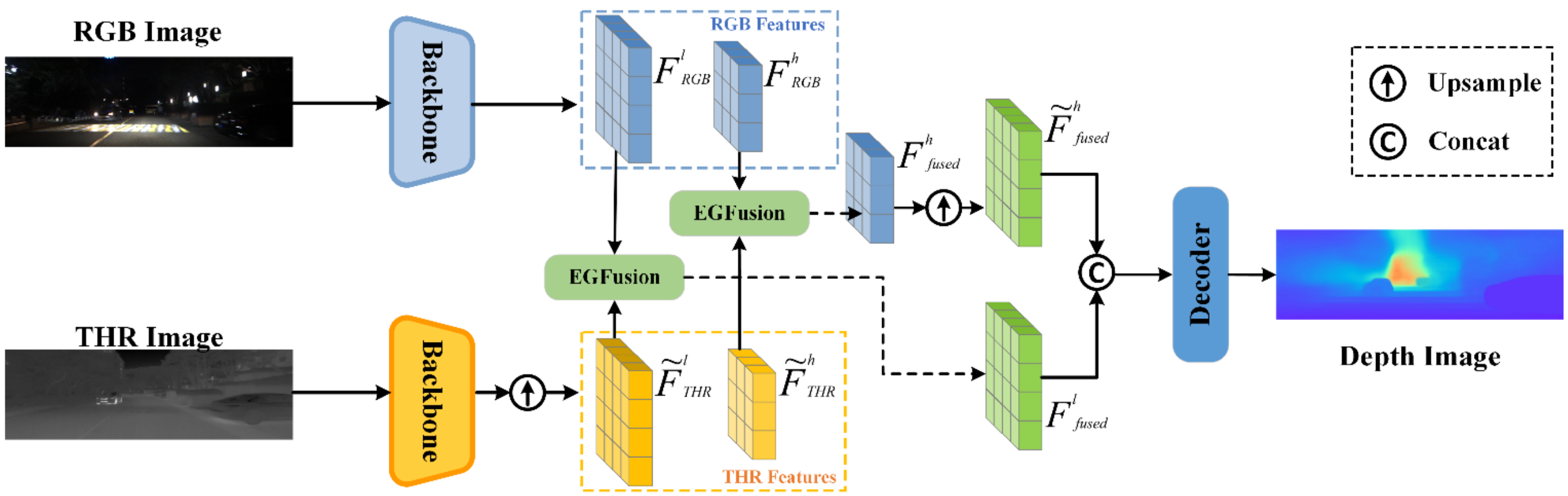

Figure 2. Overall architecture of the RTFusion depth estimation networks.

However, THR imagery lacks the detailed texture and fine-grained information available in RGB data, often leading to less precise depth maps, especially in scenes with complex geometries or low thermal contrast. As a result, THR-based methods are somehow insufficient for robust depth estimation across diverse environments.

### *C. Depth Estimation from Fused RGB and THR*

Fusing RGB and THR modalities enhances depth estimation by leveraging the strengths of both domains. RGB provides rich texture and semantic information, while THR ensures robust performance in low-visibility environments. Multi-modal fusion frameworks have been proposed to integrate dominant depth cues from both sources using confidence-based fusion networks [25]. Some methods employ cross-modality transformers to enhance feature interaction between RGB and THR, improving depth prediction accuracy in complex scenarios [26]. Additionally, self-supervised learning frameworks have been designed to exploit multi-spectral image pairs, ensuring better depth generalization across varying environments [27]. A novel approach, "MURF," integrates image registration and fusion to correct parallax errors while extracting aligned multi-modal depth representations, further improving depth estimation performance [29].

Recent approaches integrate RGB and THR data to leverage RGB's texture details and THR's thermal stability, addressing single-modality limitations. While early fusion methods concatenate features directly, they often suffer from redundancy and conflicts. Advanced strategies, such as mid-level and late fusion, use attention models, gated fusion, or feature alignment to enhance complementarity [29, 30]. However, scalability and feature alignment remain challenges. To address these issues, our method employs hierarchical multi-scale fusion with Mutual Complementary Attention (MCA) for effective RGB-THR integration and an Edge Saliency Enhancement Module (ESEM) to refine depth predictions, improving robustness and accuracy in complex environments.

## III. Proposed Methods

### *A. Model Architecture*

The proposed model employs a modular architecture tailored for effective multimodal fusion. It consists of three primary components: feature extraction, feature fusion, and decoding.

**Feature Extraction**. Independent feature extraction branches are designed for RGB and THR modalities. The RGB branch employs an optimized ConvNeXt Tiny architecture to capture detailed texture, edge, and color information. Conversely, the THR branch focuses on extracting thermal gradients and edge features specific to thermal data. Both branches incorporate a hierarchical multi-scale structure, progressively capturing fine-grained local details and global context. This modular design ensures that modality-specific characteristics are preserved, enabling optimal feature extraction for subsequent fusion.

We are given an RGB image and a THR (thermal cameras) image, and the network outputs the expected depth map. The input data is defined as follows:

$$X^{RGB} \in \mathbb{R}^{3\times H\times W}, X^{THR} \in \mathbb{R}^{1\times H\times W} \quad (1)$$

The RGB and THR images are separately fed into two ConvNeXt Tiny backbone networks with non-shared parameters.

$$F^l_{RGB}, F^h_{RGB} = ConvNeXtTiny_{RGB}(x^{RGB}) \quad (2)$$

$$F^l_{THR}, F^h_{THR} = ConvNeXtTiny_{THR}(x^{THR}) \quad (3)$$

Here, $F^l_{RGB}$ and $F^h_{RGB}$ denote the low-level and high-level feature representations extracted from the RGB modality, $F^l_{THR}, F^h_{THR}$ correspond to the low-level and high-level feature representations derived from the THR modality.

Then, we upsample the features from the THR modality to align them with those from the RGB modality and subsequently incorporate learnable positional embeddings to capture spatial dependencies.

$$\tilde{F}^l_{THR} = Interp(F^l_{THR}) \quad (4)$$

$$\tilde{F}^h_{THR} = \mathrm{Interp}(F^h_{THR}) \quad (5)$$

### *B. Feature Fusion*

To fully exploit the complementary properties of RGB and THR modalities, we propose the Edge-Guided Fusion (EGFusion) module, which consists of two key components: Mutual Complementary Attention (MCA) for cross-modal feature alignment, and Edge Saliency Enhancement Module (ESEM) for structure-aware fusion.

For the low-level feature, the process is defined as follows:

$$F^l_{fused} = EGFusion(F^l_{RGB}, F^l_{THR}) \tag{6}$$

Specifically, the EGFusion operation can be explained as the following process.

Add learnable position embeddings to capture spatial dependencies:

$$\hat{F}^l_{\mathrm{RGB}} = F^l_{\mathrm{RGB}} + \mathrm{PE}, \hat{F}^l_{\mathrm{THR}} = \tilde{F}^l_{\mathrm{THR}} + \mathrm{PE} \tag{7}$$

$$\hat{F}^h_{\mathrm{RGB}} = F^h_{\mathrm{RGB}} + \mathrm{PE}, \hat{F}^h_{\mathrm{THR}} = \tilde{F}^h_{\mathrm{THR}} + \mathrm{PE} \tag{8}$$

At this stage, the superscript ^ denotes features with position encoding applied, while $PE$ represents the position encoding itself.

**Mutual Complementary Attention (MCA).** To effectively capture the complementary relationships between RGB and THR modalities, we design a Mutual Complementary Attention (MCA) mechanism based on a Query-Key-Value (QKV) structure, enabling dynamic interaction between the two modalities. Given the input RGB and THR features, we first extract attention components and to reduce computational complexity and enlarge the receptive field, we apply a $3 \times 3$ convolution with stride $s = 2$ for downsampling:

$$Q^l = Conv_{1\times1}(\hat{F}^l_{RGB}), \bar{Q}^l = Conv^{s=2}_{3\times3}(Q^l) \tag{9}$$

$$K^l, V^l = Conv_{1\times1}(\hat{F}^l_{TIR}), \bar{K}^l, \bar{V}^l = Conv^{s=2}_{3\times3}(K^l, V^l) \tag{10}$$

Where $Q^l$ is extracted from $\hat{F}^l_{RGB}$ via a $1 \times 1$ convolution, while $K^l$, $V^l$ are obtained from $\hat{F}^l_{THR}$ using the same $1 \times 1$ convolution operation. To reduce computational complexity and enlarge the receptive field, the downsampled representations $\bar{Q}^l$, $\bar{K}^l$, $\bar{V}^l$ are obtained through a $3 \times 3$ convolution with a stride of $s = 2$.

The attention weights and output are computed:

$$A^l = softmax\left(\frac{\bar{Q}^l \bar{K}^l}{\sqrt{d_k}}\right) \tag{11}$$

Here, $\sqrt{d_k}$ serves as a scaling factor to prevent excessively large values that could lead to gradient vanishing. The attention weights $A^l$ are applied to the Value to compute the weighted sum:

$$O^l = A^l \bar{V}^l \tag{12}$$

Since the attention computation is performed on the downsampled features, we apply an upsampling operation to restore the original resolution:

$$\tilde{O}^l = Interp(O^l) \tag{13}$$

where $Interp$ denotes bilinear interpolation.

Finally, the RGB features enhanced by cross-modal attention are defined as follows:

$$F^l_{cross} = \hat{F}^l_{RGB} + \tilde{O}^l \tag{14}$$

**Edge Saliency Enhancement Module (ESEM)**. The computation process of ESEM is formulated as:

$$F^l_{\mathrm{enhanced}} = \mathrm{ESEM}(F^l_{\mathrm{THR}}, F^l_{\mathrm{cross}}) \tag{15}$$

Where the input $F^l_{\mathrm{THR}}$ is the low-scale feature extracted from the THR modality. $F^l_{\mathrm{cross}}$ is the cross-attention RGB feature obtained from MCA.

Specifically, ESEM can be defined as follows:

First, an edge weight map $E^l$ is computed by applying a two-layer convolutional network to the thermal feature map $\hat{F}^l_{\mathrm{THR}}$:

$$E^l = \sigma(\boldsymbol{W_2} \cdot ReLU(\boldsymbol{W_1} \cdot \hat{F}^l_{THR})) \tag{16}$$

$$F^l_{enhanced} = F^l_{cross} + E^l \odot \hat{F}^l_{THR} \tag{17}$$

Here, $E^l$ represents the learned edge weight map, which is generated by applying a two-layer convolutional network parameterized by $\boldsymbol{W_1}$ and $\boldsymbol{W_2}$, both of which are learnable parameters. Specifically, $\boldsymbol{W_1}$ is a 3 x 3 convolutional kernel that extracts local structural features and enhances edge-related information, while $\mathbf{W_2}$ is a 1 x 1 convolutional kernel responsible for aggregating channel-wise information and reducing dimensionality to produce a single-channel edge weight map. The output is further refined through a non-linear activation function, where ReLU enhances feature discrimination after $\boldsymbol{W_1}$, and Sigmoid normalizes the final response to the range [0,1], ensuring its suitability for edge-aware fusion. The edge-aware weighting mechanism leverages the thermal gradient information to adaptively adjust the contribution of THR features in the fused representation. High values of $E^l$ indicate strong edge presence, where thermal information is crucial for preserving object boundaries. In such regions, THR features are emphasized to enhance structural details. Conversely, lower values of $E^l$ suppress the contribution of THR information in homogeneous regions, preventing irrelevant thermal noise from affecting the RGB features. This ensures that the fusion process is guided by meaningful structural cues rather than indiscriminate feature blending.

Next, we employ convolutional layers to further integrate the cross-modal features:

$$F^l_{fused} = FusionConv(F^l_{enhanced}) \tag{18}$$

After obtaining the $F^l_{fused}$, we apply the same operations to the high-level features. Consequently, we derive the high-scale fused feature representation, denoted as $F^h_{fused}$.

After aligning the high-level and low-level features using bilinear interpolation, the feature fusion is performed:

$$F_{concat} = Concat\left(F^l_{fused}, Interp(F^h_{fused})\right) \tag{19}$$

### *C. Decoding Mechanism*

The decoding stage reconstructs high-resolution depth maps from the fused features. A multi-stage upsampling architecture progressively refines depth predictions, integrating hierarchical features from earlier stages to preserve both detail and consistency, ensuring accurate and realistic depth reconstructions. The decoder is employed to generate the final depth estimation:

$$D = UNetDecoder(F_{concat}) \tag{20}$$

where D represents the output depth map.

### *D. Loss Function*

The $L1$ loss minimizes the absolute error between the predicted depth map and the ground truth. This ensures that the overall structure of the scene is accurately captured, reducing large deviations in predicted depth values:

$$L_{L1} = \frac{1}{N}\sum \left|D_{pred}(i) - D_{gt}(i)\right| \quad (21)$$

where $N$ is the total number of pixels, and $i$ indexes the pixel locations.

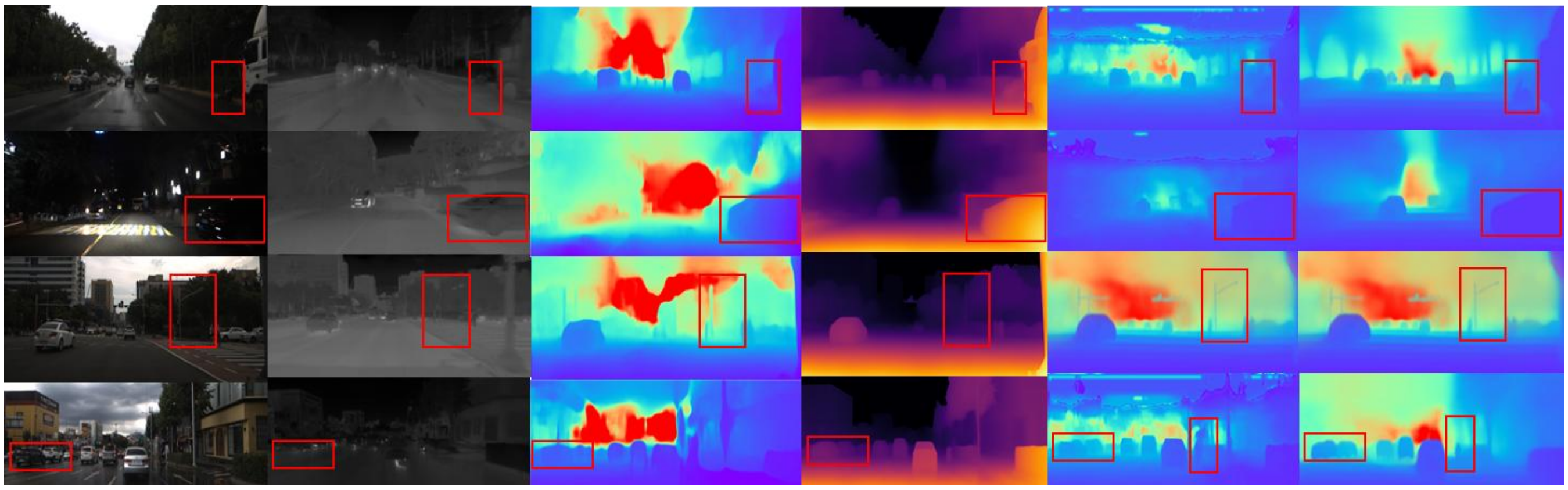

Figure 3. Qualitative results of different depth estimation methods. First column: RGB inputs. Second column: THR inputs. Third column: results from Monodepth [24]. Forth column: results from large model [12]. Fifth column: results from LapDepth [16] Sixth column: results from the proposed method RTFusion.

To enhance the accuracy of depth predictions at object boundaries, an edge-preserving smoothness term is introduced. This term penalizes large depth variations in non-edge regions while preserving sharp transitions at edges. The edge-preserving smoothness loss is defined as:

$$L_{smooth} = \sum_{i} \left|\nabla \boldsymbol{D}_{pred}(i)\right| \cdot \exp\left(-|\nabla \boldsymbol{I}(i)|\right) \quad (22)$$

where $\nabla \boldsymbol{D}_{pred}$ represents the depth gradient, $\nabla \boldsymbol{I}$ is the image intensity gradient from the RGB modality, and the exponential term reduces the penalty near high-gradient (edge) regions.

The final loss function combines the $L1$ loss and the edge-preserving smoothness loss to achieve a balance between global accuracy and boundary refinement:

$$L_{total} = \lambda_1 L_{L1} + \lambda_2 L_{smooth} \quad (23)$$

where $\lambda_1$ and $\lambda_2$ are weighting factors that control the contribution of each loss component. These parameters are tuned to ensure that the model optimizes both accuracy and edge precision effectively.

## IV. Experiments

### A. Dataset and Augmentation

The MS2 dataset (Multi-Spectral 2) is a dataset designed to contain pairs of RGB (visible light) and thermal (infrared) images, often used for research in areas such as object detection, tracking, and scene recognition under different environmental conditions. The dataset is typically utilized in multi-modal machine learning tasks, where models need to process and integrate information from both visible and thermal images.

The dataset includes synchronized RGB and thermal images captured by specialized cameras. Thermal images provide information that is not visible in the RGB spectrum, such as heat signatures from objects or individuals. The dataset includes images captured in a variety of environmental conditions and scenarios. This can include outdoor settings, indoor environments, and nighttime captures where thermal imaging provides critical information not available in RGB images. Since thermal imaging can provide unique insights, it is often used in conjunction with RGB data to enhance object detection, especially under poor lighting conditions or challenging environments (e.g., night, fog, or rain).

The proposed model was evaluated on multiple publicly available MS2 datasets to ensure comprehensive testing across diverse environmental conditions. The datasets included scenes captured during daytime, nighttime, and under challenging conditions such as fog, rain, and glare. These scenarios provided a robust testbed for evaluating the multimodal fusion approach.

The RGB and THR images were preprocessed to ensure alignment and scale consistency. Data augmentation techniques, including random cropping, flipping, and brightness adjustment, were applied to improve generalization capability of the model. The evaluation metrics used include root mean squared error (RMSE), etc.

In addition, generative AI-based style transfer techniques are utilized for data augmentation to improve model generalization.

In specific, real images are transformed into various weather or lighting conditions while preserving structural content. The original depth maps are retained as ground truth to enhance the model's robustness under diverse appearances. Secondly, structure-preserving losses such as structural similarity and edge-aware constraints are applied to maintain geometric consistency between the transformed and original images. This encourages the model to deliver accurate depth predictions despite visual variations.

TABLE 1. QUANTITATIVE EVALUATIONS OF DEPTH ESTIMATION RESULTS ON THE MS2 DATASET

| Method | | Modality | Metrics | | | | | | |
|---|---|---|---|---|---|---|---|---|---|
| | | | AbsRel↓ | SqRel↓ | RMSE↓ | RMSE(log)↓ | $a_1$ ↑ | $a_2$ ↑ | $a_3$ ↑ |
| MSFT [20] | 2022 | THR | 0.128 | 1.264 | 5.691 | 0.266 | 0.821 | 0.919 | 0.965 |
| BTS [9] | 2019 | THR | 0.129 | 1.008 | 5.169 | 0.206 | 0.843 | 0.947 | 0.978 |
| MTVUMCL [27] | 2021 | THR | 0.155 | 1.367 | 5.944 | 0.280 | 0.811 | 0.917 | 0.963 |
| MS2 (Stereo) [7] | 2023 | THR | 0.120 | 1.057 | <u>5.068</u> | 0.207 | 0.872 | 0.949 | 0.975 |
| LapDepth [16] | 2021 | RGB | 0.125 | 1.054 | 5.394 | 0.223 | 0.853 | 0.944 | 0.979 |
| MIM [23] | 2023 | RGB | 0.124 | 1.102 | 5.283 | 0.216 | 0.855 | 0.940 | 0.970 |
| AFNet [28] | 2024 | RGB | 0.122 | 1.093 | 5.214 | 0.220 | 0.866 | 0.947 | 0.973 |
| MCT [30] | 2023 | THR+RGB | <u>0.116</u> | <u>1.015</u> | 5.192 | <u>0.202</u> | <u>0.884</u> | <u>0.951</u> | <u>0.980</u> |
| MURF [29] | 2023 | THR+RGB | 0.121 | 1.062 | 5.124 | 0.210 | 0.869 | 0.948 | 0.973 |
| RTFusion (Ours) | | THR+RGB | **0.104** | **0.942** | **4.997** | **0.186** | **0.895** | **0.974** | **0.991** |

### *B. Evaluation Metrics*

The evaluation metrics used for quantitative evaluation are defined as follows:

$$AbsRel = \frac{1}{N}\sum_{i=1}^{N}\left|\frac{y_i - y_i^{pred}}{y_i}\right| \quad (24)$$

$$SqRel = \frac{1}{N}\sum_{i=1}^{N}\left(\frac{y_i - y_i^{pred}}{y_i}\right)^2 \quad (25)$$

$$RMSE = \sqrt{\frac{1}{N}\sum_{i=1}^{N}\left(y_i - y_i^{pred}\right)^2} \quad (26)$$

$$RMSE(log) = \sqrt{\frac{1}{N}\sum_{i=1}^{N}(\log((y_i+1) - \log(y_i^{pred}+1)))^2} \quad (27)$$

$$r(p) = \min\left(\frac{\hat{D}(p)}{D(p)}, \frac{D(p)}{\hat{D}(p)}\right) \quad (28)$$

### *C. Experiment Results*

In this study, we conducted extensive experiments on the MS2 dataset to evaluate the proposed RTFusion method against state-of-the-art single-modal and multimodal depth estimation approaches. The results show that our method achieves competitive performance compared to the existing techniques across multiple evaluation metrics.

Table 1 and Table 2 present the experimental results on the MS2 datasets, respectively. As shown in the results, the proposed RTFusion method outperforms other approaches in key metrics such as AbsRel, SqRel, and RMSE. Specifically, on the MS2 dataset, our method achieves an RMSE of 4.997, significantly surpassing MCT (5.192) and MURF (5.124). These findings indicate that RTFusion enables more accurate scene depth reconstruction, particularly in complex environments. In addition, as observed in Tables 1 and 2, multi-modal fusion methods based on RGB and THR consistently achieve better results than single-modal methods that rely solely on RGB or THR.

In addition to quantitative evaluations, we conducted a visual comparison as illustrated in Figure 1 and Figure 3, showing that traditional RGB-based methods struggle in low-light conditions, often producing blurred depth maps. In contrast, RTFusion effectively integrates RGB and THR data, achieving clearer and more accurate depth predictions, particularly in challenging regions. The Edge Saliency Enhancement Module (ESEM) refines depth edges by leveraging THR's temperature contrast, reducing blurring and discontinuities. Additionally, the Mutual Complementary Attention (MCA) mechanism enhances depth estimation in occluded and low-contrast areas, ensuring smoother and more reliable predictions.

### *D. Ablation Studies*

**Multimodal Fusion Strategy**. To further validate the effectiveness of multimodal fusion in the proposed RTFusion method, we conducted additional experiments on the MS2 dataset to analyze the performance of different approaches across various scenarios, including daytime, nighttime, and rainy conditions. The detailed experimental results are presented in Table 3, where RGB denotes the use of only RGB image data within the original RTFusion framework (i.e., without the THR branch), and THR represents the use of only THR image data (i.e., without the RGB branch).

As shown in Table 2, the proposed multimodal fusion depth estimation method consistently achieves the best performance across all scenarios, with particularly notable improvements in nighttime and rainy conditions. For example, in terms of the AbsRel metric, RTFusion reduces the error by 0.045 compared to the RGB single-modal approach and by 0.009 compared to the THR single-modal approach in nighttime scenarios. In rainy conditions, RTFusion achieves an error reduction of 0.008 compared to the RGB single-modal method and 0.015 compared to the THR single-modal method.

Moreover, an analysis of Table 2 shows that THR outperforms RGB by a considerable margin in nighttime scenarios, likely because thermal cameras excel at capturing scene details in low-light conditions. In conclusion, these experimental findings indicate that the proposed multimodal fusion approach provides significant benefits in addressing challenging environments, such as adverse weather and poor lighting conditions.

To evaluate the contributions of each component in the proposed model, ablation studies were performed by removing the cross-modal attention mechanism (MCA), the edge saliency enhancement module (ESEM) and analyzing their impact on depth estimation performance.

**Mutual Complementary Attention**. The MCA dynamically aligns RGB and THR features to fully leverage their complementary strengths. As shown in Table 4, when this mechanism was removed, AbsRel increased from 0.104 to 0.121, and RMSE increased to 5.226. These results demonstrate that the MCA plays a crucial role in enhancing cross-modal feature interaction, thereby improving the accuracy of depth estimation.

**Edge Saliency Enhancement Module**. The ESEM module exploits the thermal contrast inherent in THR data to improve the precision of depth predictions, particularly along object boundaries. As indicated in Table 3, the exclusion of ESEM led to an increase in RMSE from 4.997 to 5.228 and a rise in AbsRel to 0.119. Qualitative analysis of the depth maps further revealed significant boundary blurring, emphasizing the critical role of ESEM in preserving edge sharpness.

Our proposed method demonstrates competitive performancce in depth estimation, as evidenced by both comparative evaluations and ablation studies. The proposed fusion method enables effective integration of RGB and THR features, while the ESEM module enhances estimation performance by refining object boundaries. In summary, these modules form a robust and reliable network that facilitates accurate depth estimation across a wide range of challenging environments.

## V. CONCLUSIONS

The proposed multimodal fusion depth estimation model effectively leverages the complementary strengths of RGB and THR data, enabling robust and accurate depth prediction across a wide range of challenging scenarios. Its innovative architecture significantly enhances performance in complex real-world applications—such as autonomous driving and robotics—where precise depth estimation is critical. By adapting dynamically to varying environmental conditions, the model underscores the potential of multimodal fusion for real-time, high-reliability applications.